\documentclass[useAMS,usenatbib]{mn2e}

\usepackage{graphicx}

\title[2-D models of layered protoplanetary discs]{2-D models of layered protoplanetary discs: \\I. The ring instability}
\author[R.~W\"unsch, H.~Klahr and M.~R\'o\.zyczka]{R. W\"unsch$^{1,3}$\thanks{E-mail:
richard.wunsch@matfyz.cz} H.~Klahr$^{2}$ and M.~R\'o\.zyczka$^{1}$\\
$^{1}$Nicolaus Copernicus Astronomical Center, Bartycka 18, 00-716 Warsaw, Poland\\
$^{2}$Max-Planck-Institut f\"ur Astronomie, K\"onigstuhl 17, D-69117 Heidelberg, Germany\\
$^{3}$Astronomical Institute, Academy of Sciences of the Czech
Republic, Bo\v{c}n\'\i\ II 1401, 141 31 Prague, Czech Republic}
\begin{document}

\date{Received: }

\pagerange{\pageref{firstpage}--\pageref{lastpage}} \pubyear{2005}

\maketitle

\label{firstpage}

\begin{abstract} 

In this work we use the radiation hydrodynamic code TRAMP to
perform a two-dimensional axially symmetric model of the layered
disc. Using this model we follow the accumulation of mass in the
dead zone due to the radially varying accretion rate. We found a
new type of instability which causes the dead zone to split into
rings. This "ring instability" works due to the positive feedback
between the thickness of the dead zone and the mass accumulation
rate.

We give an analytical description of this instability, taking into
account non-zero thickness of the dead zone and deviations from
the Keplerian rotational velocity. The analytical model agrees
reasonably well with results of numerical simulations. Finally, we
speculate about the possible role of the ring instability in
protoplanetary discs and in the formation of planets.

\end{abstract}

\begin{keywords}
Solar system: formation, accretion discs, hydrodynamics, instabilities.
\end{keywords}

\section{Introduction} 

The layered disc model was proposed by \citet{gammie96} to account for
accretion-related phenomena in T Tauri stars. He assumed that the
angular momentum is transported by the magneto-rotational instability,
commonly referred to as the MRI \citep{bh91}. However, in the outer
disc (beyond $\sim 0.1$~AU) the temperature and the ionization degree
is so low that the gas is not well coupled to the magnetic field and
the MRI decays. There, the only parts of the disc in which the MRI can
operate are the surface layers that are ionized by cosmic rays
(ionization due to x-ray quanta emitted by the central star was also
considered; see \citealt*{gni97}). Sandwiched between the active
surface layers is an MRI-free, and, consequently, non-viscous area
near the mid-plane of the disc, commonly referred to as the dead zone.

An interesting property of layered discs is that, in general, the
accretion rate $\dot M$ is a function of the radius $r$. The specific
form of this function depends on the mass-weighted opacity. However,
$\dot M$ increases with $r$ if the opacity does not depend on the
density and increases with the temperature not faster than $T^2$. This
is true in the range $203$~K $<$ $T$ $<$ $2290\rho^{2/49}$~K, where
the opacity is dominated by ice-free grains, and also between $167$~K
and $203$~K, where the opacity drops  due to the sublimation of ices
as $T$ grows. At temperatures below $167$~K the opacity varies as
$T^2$, and the accretion rate is constant as a function of $r$ (in
this paper we use opacities of gas and dust mixture according to
\citealt{bl94}).

When $\dot M$ increases with $r$, an annulus centred at $r_0$
receives more mass per unit time from the outer disc $(r>r_0)$ than
that it loses to the inner disc $(r<r_0)$. As a result, the mass
accumulates in the dead zone. Eventually, at one or more locations in
the dead zone the surface density becomes so large that the heat
released by the accretion of the accumulated matter pushes the
temperature to a level at which the collisional ionization can restore
the coupling between the gas and the magnetic field. In such case a
triggering event, e.g. perturbation due to the passage of the
companion star, heat flux from the inner active disc or gravitational
instability of the dead zone \citep*{alp01}, could start the accretion
and "ignite" the dead zone, making at least part of it active.
Consequently, the rate at which mass is accreted by the central star
could increase dramatically. This mechanism was suggested by
\citet{gammie96} to explain the FU Ori type outbursts \citep{hk96}.

The layered disc model was further developed by \citet{hure02}, who
noted that the non-zero thickness of the dead zone is an important
factor influencing the structure of the active surface layers (this is
because their structure depends on the vertical component of gravity,
which in turn depends on the distance from the mid-plane). On the
other hand, detailed MHD simulations of surface layers performed in a
3D shearing box approximation by \citet{fs03} showed that MRI-driven,
non-axisymmetric density waves can propagate far into the dead zone.
As a result, a purely HD turbulence is excited there, providing an
effective viscosity.

Our original intention was to study how the evolution of the dead zone
may be affected by the radiative heating from the inner active part of
the disc. To that end, we performed extensive two-dimensional
(axisymmetric) simulations of the evolution of layered discs, allowing
for radiative energy transfer in both radial and vertical directions.
Unexpectedly, we found that the dead zone is unstable in a way that
has not been reported before; namely, it tends to decompose into
rings. In the present paper we illustrate this "ring" instability with
numerical simulations, and we discuss it analytically, providing a
physical explanation of the observed phenomena. The remaining results
of our simulations will be reported in a forthcoming paper.

The outline of the present paper is as follows. In \S2 we briefly
describe the numerical code and we list the assumptions underlying
our simulations. A layered-disc model whose dead zone decomposes
into rings is presented in \S3. \S4 contains an analytical
discussion of the ring instability. Finally, in \S5 we summarize
our results and discuss the effects of the ring instability in
more sophisticated disc models.


\section{Numerical methods} 

The simulations are performed with the help of the 3-D code TRAMP
\citep*{khk99}. The equations of continuity
\begin{equation}\label{ttcont}
\frac{\partial\rho}{\partial t} + \nabla\cdot(\rho\mathbf{v}) = 0 \ ,
\end{equation}
momentum conservation
\begin{equation}\label{ttmc}
\frac{\partial\rho\mathbf{v}}{\partial t} +
(\mathbf{v}\cdot\nabla)\rho\mathbf{v} =  -\nabla P +
\rho\nabla\Phi + \mathbf{f}_\mathrm{cen} + \nabla\cdot\mathbf{T} \
,
\end{equation}
and internal energy
\begin{equation}\label{ttie}
c_v\rho \left[ \frac{\partial T}{\partial t} +
(\mathbf{v}\cdot\nabla)T  \right] = -P\nabla\cdot\mathbf{v} +
\mathbf{T}:(\nabla\mathbf{v})
\end{equation}
are solved in spherical coordinates $(r,\theta,\phi)$ using an
explicit operator-splitting method. $\Phi$ is here the
gravitational potential, $\mathbf{f}_\mathrm{cen}$ is the
centrifugal force, $\mathbf{T}$ is the viscous stress tensor,
$\nabla\mathbf{v}$ is the rate of strain tensor composed of
derivatives of velocity components, $T$ is the temperature
(assumed to be the same for gas, dust and radiation), $c_v$ is the
specific heat, and the colon denotes a double scalar product of
two tensors. The self-gravity of the disc is neglected, so that
\begin{displaymath}
\Phi = \frac{GM_\star}{r} \ ,
\end{displaymath}
where $M_\star$ is the mass of the central star. We use the
equation of state of the ideal gas
\begin{equation}\label{tteos}
P = \frac{k_\mathrm{B}T}{\mu m_\mathrm{H}}\rho \ ,
\end{equation}
where $k_\mathrm{B}$ is Boltzmann constant, $\mu$ is the average
molecular weight of the disc gas, and $m_\mathrm{H}$ is the proton
mass.

The radiation transport is treated at the end of each hydrodynamic
time-step by solving the equation
\begin{equation}\label{ttrte}
\frac{\partial E_r}{\partial t} =
-\nabla\cdot\mathbf{F} \ ,
\end{equation}
where $E_r= aT^4$ is the radiation energy density, and
$\mathbf{F}$ is the radiative flux. We adopt
\begin{equation}\label{ttrflux}
\mathbf{F} = -\frac{\lambda c}{\rho\kappa}\nabla E_r \ ,
\end{equation}
where $\kappa$ is the Rosseland mean opacity for the gas-dust mixture
\citep{bl94}, and $\lambda$ is the flux limiter used to interpolate
between optically thin and optically thick cases \citep{lp81}.
Equations (\ref{ttrte}) and (\ref{ttrflux}) are solved implicitly
using an iterative SOR method. After the convergence has been
achieved, the updated temperature is calculated from the updated
radiation energy density. The advection routine for mass, momentum and
energy is based on a second-order monotonic transport scheme
originally introduced by \citet{vanleer77} and optimized by
\citet{kh87}. For more details about the numerical code we refer to
\citet{khk99}.

The models are axially symmetric. We allow for the flow through
the midplane of the disc, i.e. we do not impose a reflecting
boundary condition there. Grid points are spaced logarithmically
in $r$, resulting in a progressively smaller radial extent of grid
cells near the centre of the disc, where their vertical extent is
also progressively smaller. Thus, the shape of grid cells is more
nearly regular throughout the grid.

In the inner active part of the disc, approximate initial
distributions of density and temperature are obtained from
analytical $\alpha$-disc models \citep{ss73} assuming
a constant temperature profile in the direction perpendicular to
the midplane. The outer, layered part of the disc is initiated
with the analytical model of \citet{gammie96}. The two solutions
merge at the radius $r_{\mathrm DZ}$, where the mid-plane
temperature of the layered part reaches $1000$~K. The accretion
rate of the layered part at $r_{\mathrm DZ}$ defines the accretion
rate in the inner active part. Similarly, the surface density of
the dead zone  is calculated assuming continuity of the total
surface density at $r_{\mathrm DZ}$. Initially, the surface
density of the dead zone is constant in $r$. The initial
rotational velocity is chosen so as to balance the gravity reduced
due to the radial pressure gradient (in a thin disc the rotation
is nearly Keplerian). Other components of velocity are set to
zero.

At the outer edge of the disc mass is injected at every time-step
at a fixed rate by the following procedure:
\begin{enumerate}
\item densities and temperatures are copied into the ghost zone from
the adjacent active cells
\item densities in the ghost zone are
normalized to the initial surface density
\item a uniform radial
velocity is set across the ghost zone such that the accretion rate
is the same as in the analytical model.
\end{enumerate}
The angular rotational velocity at both inner and outer boundary
is extrapolated using a $r^{-3/2}$ power-law. At the inner edge of
the disc, and at the upper and lower boundary of the grid, a free
outflow boundary condition is imposed. A constant temperature of
$10$~K is maintained at the upper and lower boundary.

At each time-step the location of the dead zone is found based on
two conditions that have to be fulfilled simultaneously:
\begin{enumerate}
\item $T < T_\mathrm{lim}$, where $T_\mathrm{lim}$ is the minimum
temperature at which the coupling between the gas and the magnetic
field still operates
\item$\Sigma > \Sigma_a$, where $\Sigma$ is the column density
integrated from the surface of the disc along a line perpendicular
to the midplane, and $\Sigma_a$ is the column density ionized by
the cosmic rays.
\end{enumerate}
Beyond the dead zone the viscosity coefficient is defined
according to the $\alpha$-prescription of \citet{ss73}
\begin{equation}\label{s-s}
\nu = \alpha c_\mathrm{s} H_a \ ,
\end{equation}
where $\alpha$ is a dimensionless parameter, $c_s$ is the sound
speed and $H_a$ is the thickness of the active layer which is
determined from the vertical distribution of density for each
radius at each time-step. We chose this particular form of
Shakura-Sunyaev prescription because it is closer to the analytical,
vertically averaged model. Within the dead zone we set $\nu=0$.

To avoid numerical problems, we introduce a density limiter:
whenever the density falls below $\rho_{min}$ (which may be
different for different models), it is doubled, and the
temperature is adjusted so as to keep the pressure unchanged. This
procedure leads to the formation of a low-density "atmosphere"
surrounding the disc. To prevent it from collapsing onto the disc,
we artificially damp vertical and radial velocities in this
region. However, this artificial readjustment only affects a
negligible number of grid cells in the "atmosphere", and the
overall evolution of the disc is not influenced.


\section{Results of simulations}   
\label{ldsim}

We obtained a broad sample of models with different parameters, which
will be presented elsewhere. Here we describe a representative model
with "canonical" parameters $\mu=2.353$, $T_\mathrm{lim} = 1000$~K,
$\Sigma_a = 100~\mathrm{g\,cm}^{-2}$ and $\alpha = 0.01$. Note that at
$T = 1000$~K the ionization degree increases by five orders of
magnitude due to ionization of potassium \citep{umebayashi83}, and the
magnetic Reynolds number, which is directly proportional to the
ionization degree, exceeds unity. $\Sigma = 100~\mathrm{g\,cm}^{-2}$
is the standard stopping column density of cosmic rays \citep{un81},
and 0.01 is a value of the viscosity parameter often
adopted for protoplanetary discs (e.g. \citealt*{hgb95}). The
simulations were performed on a grid of $256\times45$ points in
($r$,$\theta$), with $r$ varying between 0.05 and 0.35 AU and $\theta$
varying between $-5\degr$ and $+5\degr$. The model was integrated for
$124$~yr, i.e. for 600 orbits at the outer edge of the disc.

\begin{figure*}
\centering
\begin{minipage}{140mm}
\includegraphics{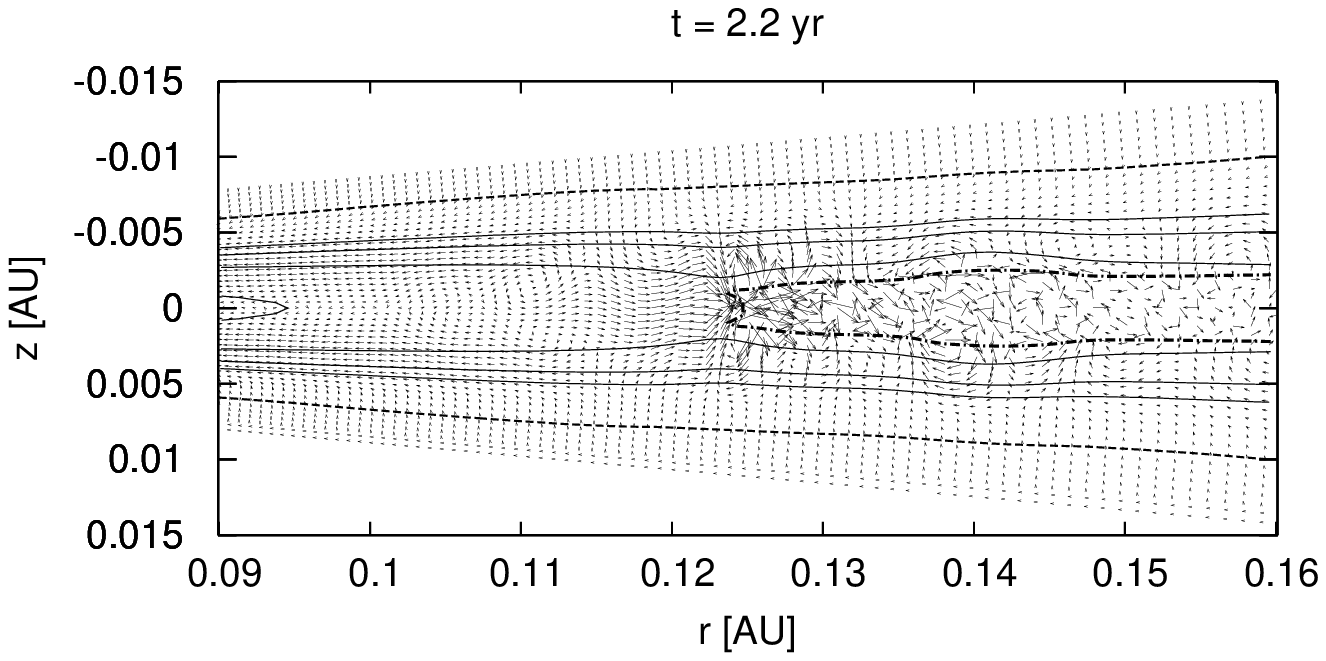}
\includegraphics{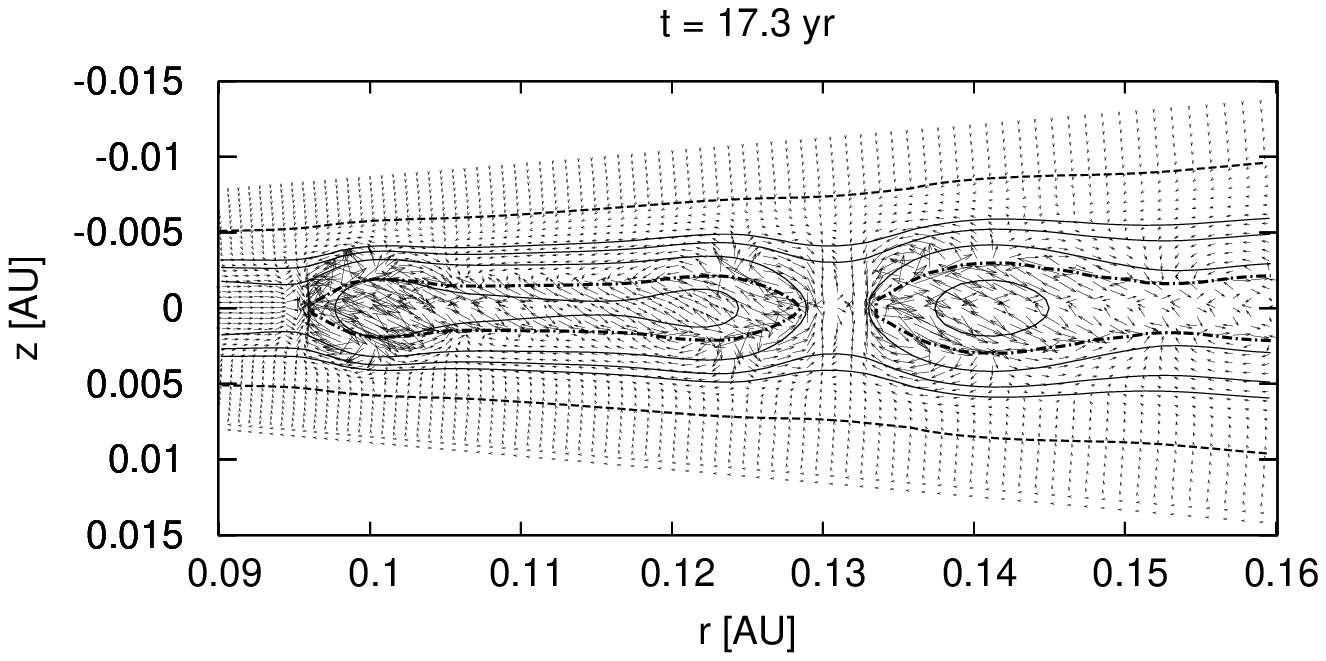}
\includegraphics{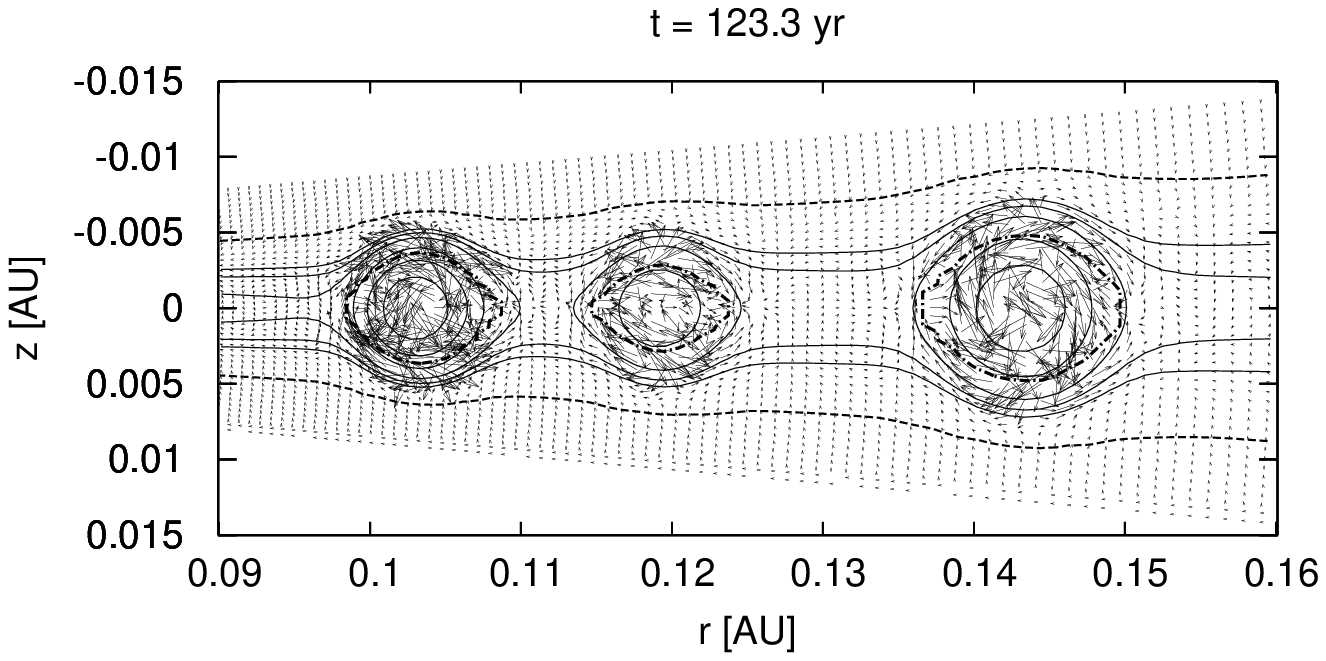}
\caption{A detail of the computational domain around forming rings
at the end of the simulation. Contours indicate density (the
levels are $5\times10^{-10}$, $10^{-9}$, $2\times10^{-9}$,
$5\times10^{-9}$, $10^{-8}$, $2\times10^{-8}$ and
$5\times10^{-8}$~g\,cm$^{-3}$), arrows denote the mass flux (for
the sake of lucidity their sizes are proportional to the square
root of the mass flux). The thick dash-dotted line marks the dead
zone. The surface of the disc (i.e. the boundary between the disc
and the atmosphere) is marked by the dashed line.} \label{rtmfdz}
\end{minipage}
\end{figure*}

The initial model is far from thermal equilibrium, however it
quickly relaxes due radiative heat diffusion. The density and mass
flux across the computational domain shortly after the relaxation,
at the beginning of the instability and at the end of the
simulation are shown in Fig.~\ref{rtmfdz}: one clearly sees how
the dead zone successively breaks into more and more rings.

\begin{figure}
\noindent
\includegraphics{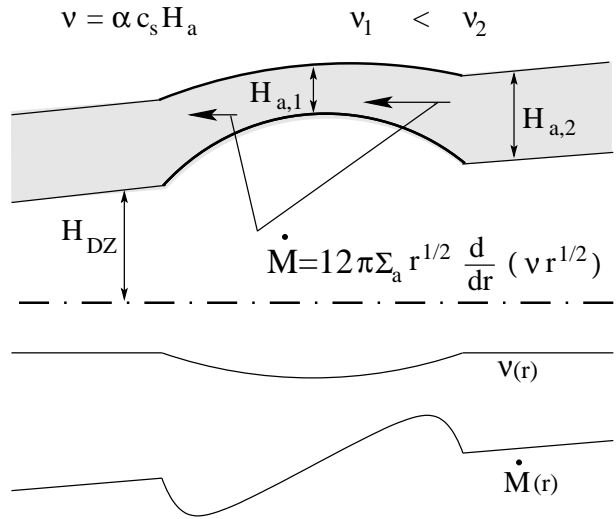}
\caption{Upper part of the figure shows the layered disc structure
with the ring-like perturbation. The higher gravitational force
compresses the elevated part of the active layer, making $H_{a,1}$
smaller than the unperturbed value $H_{a,2}$. Therefore also
$\nu_{1} < \nu_{2}$. The radial profiles of average viscosity and
 accretion rate are shown at the bottom of the
figure.} \label{ringmech}
\end{figure}

As we have not imposed any explicit perturbations, there is a good
reason to believe that the rings result from a linear instability
of layered discs. The mechanism of this instability is illustrated
in Fig.~\ref{ringmech}. Assume a small axially symmetric
enhancement of the surface density at a radius $r_0$ (which, of
course, must be accompanied by an increase of the dead zone
thickness $H_\mathrm{DZ}$). Since at higher distances from the
mid-plane the vertical component of gravity is stronger, the
active layer (whose column density is at all times fixed by cosmic
rays) must get thinner at $r_0$.

In the standard prescription, the viscosity coefficient is
proportional to the local thickness of the active zone $H_a$ (this
is motivated by the idea that $H_a$ is a natural scale that limits
the maximum size of the edies of the MHD turbulence). Therefore,
smaller $H_a$ results in a lower viscosity in the elevated part of
the active layer. The accretion rate, which depends on the
derivative of viscosity (see Eq.~\ref{efar} below), decreases at
the inner edge of the ring. This causes a bottle-neck effect in
the accretion flow, and the mass accumulates in the ring. On the
other hand, the higher accretion rate at the outer edge of the
ring makes the ring more compact. This positive feedback between
the dead zone thickness and the mass accumulation rate leads to
the formation of the rings.

\begin{figure}
\noindent
\includegraphics{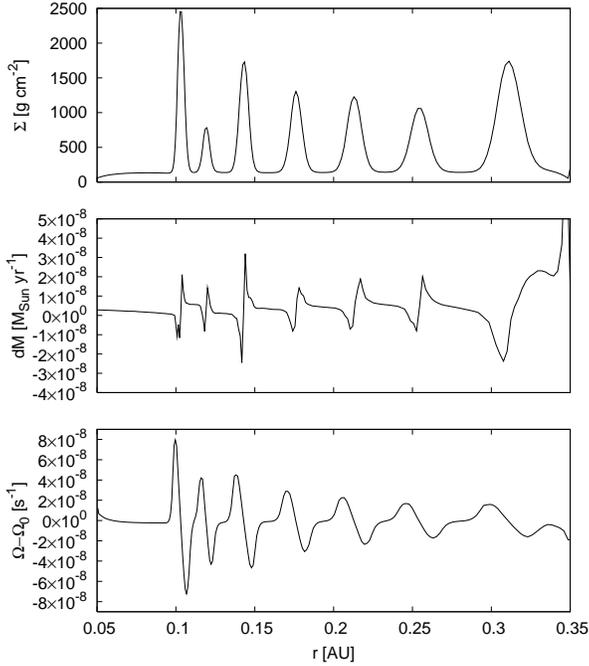}
\caption{Radial profiles of surface density $\Sigma$ (top), the accretion
rate $\dot{M}$ (middle) and difference to the Keplerian angular velocity
$\Omega - \Omega_0$ (bottom). } \label{rp-om}
\end{figure}

The rings formed in the simulation are shown in the radial profile
of the surface density in Fig.~\ref{rp-om}. The accretion rate
profile exhibits the described minima at the inner edges of the
rings and maxima at the outer ones. The plot also shows the
deviations from the Keplerian angular velocity -- it can be seen
that the inner edges of the rings rotate at a super-Keplerian
velocity. Thus, the rings may capture and concentrate the radially
drifting boulders of meter-size, preventing them from accretion
onto the central star (e.g. \citealt{kl00}). Such concentration of
solids in peaks of gas density was observed by \citet{hb03} and
\citet{rlpab04} in the case of spiral density waves formed by the
gravitational instability.


\section{Analytical description of the ring instability} 

\subsection{Basic assumptions and definitions} 

Let us consider a layered disc consisting of a dead zone with
thickness $2H_\mathrm{DZ}$ and two active surface layers with
thickness $H_a$ each. The surface layer accretes at a rate
\begin{equation}
\dot{M} = 6\pi r^{1/2} \frac{\partial }{\partial r} (2\Sigma_a \nu
r^{1/2}),
\label{efar}
\end{equation}
where $\Sigma_a$ is the column density of the surface layer, $\nu$
is the kinematic viscosity, and $r$ is the cylindrical radial
coordinate. Equation (\ref{efar}) is a direct consequence of
angular momentum conservation \citep{gammie96}. Assuming that the
accretion energy is radiated locally, we obtain the standard
formula for the effective temperature $T_e$
\begin{equation}
\frac{9}{4}\Sigma_a \nu \Omega^2 = \sigma T_e^4,
\label{ldis}
\end{equation}
where $\sigma$ is Stefan-Boltzmann constant. In the optically
thick approximation the temperature $T_i$ at the boundary between
the dead zone and the active layer is given by the formula
\begin{equation}
T_i^4 = \frac{3}{8} \tau T_e^4 \ ,
\label{hubeny}
\end{equation}
where $\tau = \Sigma_a \kappa$ is the optical depth of the active
layer, and $\kappa$ is the Rosseland mean opacity \citep{hubeny90}.
In general, the opacity coefficient is a complex function of
density and temperature. In a protoplanetary disc it can be
approximated by a set of power laws
\begin{equation}
\kappa(\rho, T) = \kappa_0 \rho^a T^b\,
\label{gen_opac}
\end{equation}
where the constants $\kappa_0$, $a$ and $b$ have different values
in different opacity regimes \citep{bl94}. At temperatures
lower than $2290\rho^{2/49}$~K (i.e. nearly everywhere in a
protoplanetary disc) $\kappa$ practically does not depend on
density, so we assume $a=0$. As before, we assume that the
viscosity coefficient is given by equation (\ref{s-s}).

Let $\delta$ be the ratio of the total disc
half-thickness to the active layer thickness $H_a$:
\begin{equation}
\delta \equiv \frac{H_a + H_\mathrm{DZ}}{H_a} \ .
\end{equation}
From the equation of hydrostatic equilibrium in the direction
perpendicular do the midplane of the disc we get an approximate
formula
\begin{equation}
\frac{P_i}{\rho_i} = \Omega^2 \delta H_a^2 = c_{si}^2 \ ,
\label{vedz}
\end{equation}
where $P_i$, $\rho_i$ and $c_\mathrm{si}$ is, respectively,
pressure, density and sound speed at the boundary between the dead
zone and the active layer. Combining (\ref{vedz}) with equations
(\ref{ldis}) and (\ref{hubeny}), and using (\ref{s-s}) with
$c_s = c_\mathrm{si}$ we get
\begin{equation}
T_i = \left[ \frac{3\kappa_0}{8} \frac{9}{4\sigma} \frac{k_B}{\mu
m_\mathrm{H}} \right]^\frac{1}{3-b} \Sigma_a^\frac{2}{3-b}
\alpha^\frac{1}{3-b} \delta^{-\frac{1}{2(3-b)}}
\Omega^\frac{1}{3-b} \ .
\label{Ti}
\end{equation}
Inserting equations (\ref{ldis}) -- (\ref{Ti}) into equation
(\ref{efar}), we arrive at the following formula for the accretion
rate in the active layer:
\begin{equation}
\dot{M} = \mathcal{M} r^{1/2} \frac{\partial }{\partial r} \left(
\Omega^\frac{-2+b}{3-b} \delta^\frac{-4+b}{2(3-b)} r^{1/2}\right)
\ ,
\label{ardo}
\end{equation}
where
\begin{equation}
\mathcal{M} = 12\pi \left[ \frac{3\kappa_0}{8} \frac{9}{4\sigma} \right]^\frac{1}{3-b}
\frac{k_B}{\mu m_\mathrm{H}}^\frac{4-b}{3-b} \Sigma_a^\frac{5-b}{3-b} \alpha^\frac{4-b}{3-b} \ .
\end{equation}
In the following, we shall derive an equation which relates the
angular rotational velocity $\Omega$ to the thickness of the disc.

Let us assume that the disc is thin ($\delta H_a \ll r$), and
neglect the dependence of $\Omega$ on $z$ (see e.g. \citealt{urpin83}). We may write
\begin{equation}
\Omega^2 = \Omega_0^2 + \frac{1}{r\rho_m} \frac{\partial
P_m}{\partial r} \ ,
\label{omper_bas}
\end{equation}
where  $P_m = c_\mathrm{si}^2\rho_m$, $\rho_m$ and $c_\mathrm{si}$
are, respectively, mid-plane values of pressure, density, and
sound speed. Note that because there is no heat generation in the
dead zone, the mid-plane sound speed is the same as at the
boundary between the dead zone and the active layer (i.e. at each
$r$ the dead zone is isothermal along a line perpendicular to the
midplane).

\noindent The mid-plane density $\rho_m$ is given by the vertical
hydrostatic equilibrium
\begin{equation}
\rho_m  = \rho_i \exp\left( \frac{\Omega^2
H_\mathrm{DZ}^2}{2c_\mathrm{si}^2} \right) = \rho_i
\exp\left(\frac{(\delta-1)^2}{2\delta}\right),
\label{rhomp1}
\end{equation}
and for $\delta\gg 1$ we have
\begin{equation}
\rho_m=\rho_i \exp\left( \frac{\delta}{2}-1 \right)
\label{rhomp2}
\end{equation}
where it was assumed that the mass accumulated in dead zone is
already large (i.e. $\delta \gg 1$), which allows us to neglect
terms of the second order in $1/\delta$. Inserting
$\rho_\mathrm{m}$ and $P_\mathrm{m}$ into equation
(\ref{omper_bas}) we obtain
\begin{equation}
\Omega^2 = \Omega_0^2 +
    \frac{c_\mathrm{si}^2}{2r}\frac{\partial\delta}{\partial r} +
    \frac{1}{r}\frac{\partial c_\mathrm{si}^2}{\partial r} +
    \frac{c_\mathrm{si}^2}{\rho_i r}\frac{\partial \rho_i}{\partial r}
    \ .
\label{omd_full}
\end{equation}
The sound speed $c_\mathrm{si}$ only weakly depends on disc
thickness ($c_\mathrm{si}^2 \sim \delta^{-1/5}$ for $b=0.5$). The
dependence of $\rho_i$ on $\delta$ is even weaker ($\rho_i \sim
\delta^{1/10}$ for $b = 0.5$). Therefore, we assume that for a
small perturbation of $\delta$ the last two rhs terms in equation
(\ref{omd_full}), which are proportional to derivatives of
$c_\mathrm{si}^2$ and $\log(\rho_i)$, are small compared to the
second rhs term, which is directly proportional to the derivative
of $\delta$.

The final relation between $\Omega$ and $\delta$ is
\begin{equation}
\Omega^2 = \Omega_0^2 +
\frac{c_\mathrm{si}^2}{2r}\frac{\partial\delta}{\partial r} \ .
\label{omd}
\end{equation}
Equations (\ref{ardo}) and (\ref{omd}) will be used in the next
subsections to derive the dispersion relation for the
perturbations of the disc.

\subsection{The linear analysis} 

Let us perturb $\delta$ at a radius $r_0$, assuming that in a
small region $(r_0-R, r_0+R)$, $R \ll r$ $\delta$ consists of the
unperturbed part $\delta_0$ (which in that region may be regarded
as independent of $r$) plus a cosine term with a wavenumber $k$
\begin{equation}
\delta = \delta_0 + \delta_k \cos(kR) \ ,
\end{equation}
where $\delta_k$ is the amplitude of the perturbation.

\noindent According to (\ref{omd}), the angular velocity consists
of the Keplerian part $\Omega_0$ and the pressure-correction
$\Omega_1$. Neglecting terms of the second order in $\Omega_1$,
and using equation (\ref{omd}) we obtain
\begin{equation}
\Omega^2 = \Omega_0^2 + 2\Omega_0\Omega_1 \quad\rightarrow\quad
\Omega_1 = \frac{c_\mathrm{si}^2}{4r\Omega_0}\frac{\partial
\delta}{\partial r} \ .
\end{equation}

\noindent The first and second derivatives of $\delta$ and
$\Omega$ are
\begin{equation}
\begin{array}{ll}
\frac{\partial \delta}{\partial R} = -\delta_k k \sin(kR) &
\frac{\partial \Omega}{\partial R} = -\frac{3\Omega_0}{2r} -
\frac{c_\mathrm{si}^2}{4r\Omega_0}\delta_k k^2 \cos(kR)
\\
\frac{\partial^2 \delta}{\partial R^2} = -\delta_k k^2 \cos(kR) &
\frac{\partial^2 \Omega}{\partial R^2} = \frac{15\Omega_0}{4r^2}
+ \frac{c_\mathrm{si}^2}{4r\Omega_0} \delta_k k^3 \sin(kR)
\end{array}
\end{equation}

\noindent Since we only want to obtain a rough idea about the
growth rate of the perturbation with a specific $k$, we may
approximate $\cos(kR)$ with 1 and $\sin(kR)$ with 0. Then the disc
thickness $\delta$, the angular velocity $\Omega$ and their
derivatives are:
\begin{equation}\label{delta_pert}
\delta = \delta_0 + \delta_k,
\quad
\frac{\partial \delta}{\partial R} = 0,
\quad
\frac{\partial^2 \delta}{\partial R^2} = - \delta_k k^2
\end{equation}
\begin{equation}\label{omega_pert}
\Omega = \Omega_0,
\quad
\frac{\partial \Omega}{\partial R} = -\frac{3\Omega_0}{2r_0} - \frac{c_\mathrm{si}^2}{4r\Omega_0} \delta_k k^2,
\quad
\frac{\partial^2 \Omega}{\partial R^2} = -\frac{15\Omega_0}{4r_0} \ .
\end{equation}

\noindent Due to mass accumulation, the surface density of the
dead zone $\dot{\Sigma}_\mathrm{DZ}$ grows at a rate
\begin{equation}\label{dSdM}
\dot{\Sigma}_\mathrm{DZ} = \frac{1}{2\pi r} \frac{\partial \dot{M}}{\partial r} \ .
\end{equation}
Inserting equation (\ref{ardo}) into (\ref{dSdM}) and using
approximations (\ref{delta_pert}) and (\ref{omega_pert}) we obtain
\begin{eqnarray}
\dot{\Sigma}_\mathrm{DZ} & = & \frac{\mathcal{M}}{2\pi r_0} \left[
\frac{3}{2}\frac{-2+b}{3-b}\Omega_0^\frac{-5+2b}{3-b}\delta^\frac{-4+b}{2(3-b)}\frac{\partial \Omega}{\partial R}
\right. \nonumber\\
& + & \left. \frac{-4+b}{2(3-b)}r_0\Omega_0^\frac{-2+b}{3-b}\delta^\frac{-10+3b}{2(3-b)}\frac{\partial^2 \delta}{\partial R^2}
\right. \nonumber\\
& + & \left. \frac{-2+b}{3-b}\frac{-5+2b}{2(3-b)}r_0\Omega_0^\frac{-8+3b}{3-b}\delta^\frac{-4+b}{2(3-b)}
\left(\frac{\partial\Omega}{\partial R}\right)^2
\right. \nonumber\\
& + & \left. \frac{-2+b}{3-b}r_0\Omega_0^\frac{-5+2b}{3-b}\delta^\frac{-4+b}{2(3-b)}\frac{\partial^2 \Omega}{\partial R^2}
\right]
\end{eqnarray}

\noindent The linearized unperturbed part of the last equation is
\begin{equation}\label{sdz0}
\dot{\Sigma}_\mathrm{DZ,0} = \frac{3\mathcal{M}}{8\pi r_0^2}\frac{-2+b}{3-b}\frac{-9+4b}{3-b}
\Omega_0^\frac{-2+b}{3-b}\delta_0^\frac{-4+b}{2(3-b)} \ ,
\end{equation}
and the linearized equation which describes the evolution of
surface density perturbation with wavenumber $k$ is
\begin{eqnarray}\label{sdzk}
\dot{\Sigma}_{\mathrm{DZ},k} & = & \frac{\mathcal{M}}{2\pi r_0^2}\delta_k
\nonumber\\
& \times & \left[
\frac{3}{4}\frac{-2+b}{3-b}\frac{-4+b}{2(3-b)}\frac{-9+4b}{3-b}\Omega_0^\frac{-2+b}{3-b}\delta_0^\frac{-10+3b}{2(3-b)}
\right. \nonumber\\
& + & \left.\frac{3c_\mathrm{si}^2}{4}\frac{-2+b}{3-b}\frac{7-3b}{2(3-b)}\Omega_0^\frac{-8+3b}{3-b}\delta_0^\frac{-4+b}{2(3-b)}k^2
\right. \nonumber\\
& - & \left. \frac{-4+b}{2(3-b)}r_0^2\Omega_0^\frac{-2+b}{3-b}\delta_0^\frac{-10+3b}{2(3-b)}k^2
\right] \ .
\end{eqnarray}

\noindent The surface density of the dead zone
$\Sigma_\mathrm{DZ}$ is related to the disc thickness $\delta$
through
\begin{eqnarray}
\Sigma_\mathrm{DZ} & = & 2\int_0^{H_\mathrm{DZ}} \rho(z) dz =
2\int_0^{H_\mathrm{DZ}} \rho_\mathrm{m} \exp\left(-\frac{\Omega^2 z^2}{2c_\mathrm{si}^2}\right) dz
\nonumber\\
& = &
2\int_0^{H_\mathrm{DZ}} \rho_i \exp\left( \frac{\Omega^2H_\mathrm{DZ}^2}{2c_\mathrm{si}^2} \right)
\exp\left(-\frac{\Omega^2 z^2}{2c_\mathrm{si}^2}\right)dz
\end{eqnarray}
where we used the mid-plane density given by (\ref{rhomp2}). The
integral can be evaluated using an error function to yield
\begin{equation}
\Sigma_\mathrm{DZ} =
\frac{\sqrt{2\pi}c_\mathrm{si}}{\Omega}\rho_i\exp\left(\frac{\delta}{2}-1\right)
\mathrm{erf}\left(\frac{\delta-1}{\sqrt{2\delta}} \right) \ .
\end{equation}
\noindent Remembering that $\delta \gg 1$ we get
\begin{equation}\label{sdzd}
\Sigma_\mathrm{DZ} = \Sigma_a\sqrt{\delta}\exp\left(\frac{\delta}{2}-1\right)
\end{equation}
%

\noindent Differentiating the last equation with respect to time
leads to a formula connecting $\dot{\Sigma}_\mathrm{DZ}$ with
$\dot\delta$
\begin{equation}\label{dsdzd}
\dot\Sigma_\mathrm{DZ} =
\Sigma_a\frac{\dot\delta\sqrt{\delta}}{2}\left(\frac{1}{\delta}+1\right)
\exp\left(\frac{\delta}{2}-1\right) \,
\end{equation}

\noindent whose unperturbed part and equation (\ref{sdz0}) can be
combined into an equation
\begin{eqnarray}
\dot\delta_0 & = & \frac{3\mathcal{M}}{4\pi
r_0^2\Sigma_a}\frac{-2+b}{3-b}
\frac{-9+4b}{3-b}\Omega_0^\frac{-2+b}{3-b}\delta_0^\frac{-7+2b}{2(3-b)}
\nonumber\\
& & \exp\left(1-\frac{\delta_0}{2}\right)\left(1+\frac{1}{\delta_0}\right)^{-1}
\ ,
\end{eqnarray}

\noindent which describes the evolution of the unperturbed disc
thickness due to the accumulation of mass. Next, from equation
(\ref{sdzk}), and equation (\ref{dsdzd}) written for a specific
wavenumber $k$ we get the dispersion relation which describes the
growth rate of the perturbation with the wavenumber $k$.
\begin{eqnarray}\label{disprel}
\dot\delta_k & = & \frac{\mathcal{M}}{\pi r_0^2\Sigma_a}
\exp\left(1-\frac{\delta_0}{2}\right)\left(1+\frac{1}{\delta_0}\right)^{-1}\delta_k
\nonumber\\
& \times & \left[ \frac{3}{4}\frac{-2+b}{3-b}
\frac{-9+4b}{3-b}\Omega_0^\frac{-2+b}{3-b}\delta_0^\frac{-7+2b}{2(3-b)}
\right.\nonumber\\
& \times & \left. \left(-\frac{\delta_0+2}{2(\delta_0+1)}+\frac{-4+b}{2(3-b)}\frac{1}{\delta_0}\right)
\right. \nonumber\\
&+&\left.\frac{3c_\mathrm{si}^2}{4}\frac{-2+b}{3-b}
\frac{7-3b}{2(3-b)}\Omega_0^\frac{-8+3b}{3-b}\delta_0^\frac{-7+2b}{2(3-b)}k^2
\right. \nonumber\\
&-&\left. \frac{-4+b}{2(3-b)}r_0^2\Omega_0^\frac{-2+b}
{3-b}\delta_0^\frac{-13+4b}{2(3-b)}k^2 \right]
\end{eqnarray}
The perturbation growth rate diverges at short wavelengths.
This is because our simple analysis does not include any damping
mechanisms. In reality, however, thin rings would diffuse due to
the thermal motion of particles.

Note also that that the radial pressure scale height cannot become
smaller than the vertical one. Therefore, rings with circular r-z
profiles should form, which is in agreement with our numerical
model (see Fig.~\ref{rtmfdz}).

\noindent
\begin{figure}
\noindent
\includegraphics{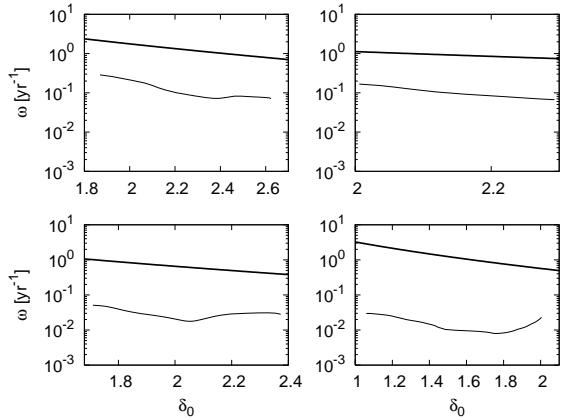}
\caption{Growth rates of the four inner-most rings
depending on the unperturbed disc thickness $\delta_0$ in comparison to growth
rates calculated from equation (\ref{disprel}). Growth rates in
the numerical simulation were determined by measuring $\delta$ and
$\dot{\delta}$ ($\omega = \dot\delta/\delta$).} \label{Od-dr1}
\end{figure}

Fig.~\ref{Od-dr1} shows growth rates of the disc thickness $\delta$ in
the four inner-most rings. They are compared to the growth rates given
by the dispersion relation (\ref{disprel}). The growth rates in the
analytical model are more than one order of magnitude higher. The
possible explanation of this discrepancy can be that the analytical
model does not allow for heat transfer from the ring into the active
layer above it or for convective flows that mix the mass inside the
rings. The radiative transfer in the radial direction may also be
important. All those processes make the dip in the average viscosity
shallower and effectively decrease the growth rate of the instability.

%
%

\subsection{The influence of stellar irradiation}

In this section we estimate how strongly the ring instability is
affected by the radiation flux from the star. In the preceding
section the stellar flux was not included self-consistently since
the linear analysis presented there serves as the comparison model
for the numerical simulation which does not include irradiation.
On the other hand, with the irradiation included explicitly the
calculations would become too complex (or just impossible) to
perform.

Firstly, we estimate how important the irradiation is for the
unperturbed disc. We do it by computing the change of the
temperature at the boundary between the dead zone and the surface
layer (from which the thickness of the surface layer and the
viscosity are determined).

Stellar heating can be included in Eq.~(\ref{hubeny}) in the
following way

\begin{equation}
T_i^4 = \frac{3}{8}\tau T_e^4 + W T_{\star}^4 \ ,
\label{irr_hubeny}
\end{equation}
where $T_e$ is the effective temperature resulting {\em solely}
from the viscous dissipation, $T_\star$ is the temperature of the
stellar photosphere, and $W$ is the dilution factor which depends
on stellar radius, distance from the star, and geometry of the
disc \citep{hubeny90}. For a conical disc (in which the aspect
ratio $H/r$ does not depend on $r$) and  $r \gg R_\star$ we have

\begin{equation}
W = \frac{2}{3\pi} \left( \frac{R_\star}{r}\right)^{3}
\end{equation}
(see e.g. \citealt{rp91}).

Evaluating the rhs terms in Eq.~(\ref{irr_hubeny}) for our disc
model and typical parameters of a T Tauri star ($R_\star =
3$~R$_\odot$, $T_\star = 4400$~K) we find that the first (viscous)
term always dominates. More susceptible to effects of irradiation
are flaring discs, in which $H/r \sim r^\gamma$. The flaring index
$\gamma$ depends on the model, e.g. a vertically isothermal model
in which the intercepted stellar flux is equal to the blackbody
emission from the disc yields $\gamma = 2/7$ \citep{cg97}.
However, even in such model the irradiation term dominates only
for ($r\ga 10$~AU), where the grazing angle becomes sufficiently
large.

Effects of irradiation may be more important for the ring
instability itself. The inner edge of a growing ring is more
strongly heated by stellar radiation, because the grazing angle
$\alpha_\mathrm{gr}$ is larger there. As a result, the temperature
and the thickness of the active layer increase locally, leading to
an increased viscosity. Then, the dip in the accretion rate, which
is responsible for the ring growth, becomes shallower or it may
even disappear entirely.

The importance of this effect may be roughly estimated by
comparing the amplitude of viscosity increase due to stellar
irradiation to the amplitude of viscosity decrease due to the
stronger vertical component of the gravitational force (the latter
effect is explained in Fig. \ref{ringmech}).

Let us assume a ring-like perturbation of the disc thickness with
an amplitude $\delta_k$ and a wavelength $\lambda = 2\pi/k$. Since
the most unstable wavelength is $\lambda_\mathrm{max} \sim
\delta_0+\delta_k$, let us parametrize the perturbation wavelength
by the dimensionless value $l = \lambda / \lambda_\mathrm{max}$.
The grazing angle at which the stellar radiation strikes the inner
edge of the ring is

\begin{equation}
\alpha_\mathrm{gr} \sim \frac{\delta_k}{l(\delta_0+\delta_k)}
+ \frac{2}{3\pi}\frac{R_\star}{r} \ .
\end{equation}
Evaluating the dilution factor $W = \alpha_\mathrm{gr}
(R_\star/r)^2$, inserting it into Eq.~(\ref{irr_hubeny}), and combining
it with Eqs.~(\ref{ldis}), (\ref{vedz}) and (\ref{s-s}) we obtain the
viscosity in a form

\begin{eqnarray}
\nu(\delta_k,\delta_0,r,l) & = &
N_1(r)(\delta_0+\delta_k)^{\frac{-4+b}{2(3-b)}}
+ N_2(r)
\nonumber\\
& \times &
(\delta_0+\delta_k)^{-1/2}
\left[ \frac{\delta_k}{l(\delta_0+\delta_k)} + N_3(r)\right]^{1/4}
\end{eqnarray}
where
\begin{equation}
N_1(r) = \left(\frac{3\kappa_0}{8}\frac{9}{4\sigma}\right)^{\frac{1}{3-b}}\alpha^{\frac{4-b}{3-b}}
\left(\frac{k_B}{\mu m_\mathrm{H}}\right)^{\frac{4-b}{3-b}}\Omega^{\frac{-2+b}{3-b}} \ ,
\end{equation}
\begin{equation}
N_2(r) = \frac{\alpha}{\Omega} \frac{k_B}{\mu m_\mathrm{H}}
\left(\frac{R_\star}{r}\right)^{-1/2} T_\star \ ,
\end{equation}
and
\begin{equation}
N_3(r) = \frac{2}{3\pi} \frac{R_\star}{r} \ .
\end{equation}

\begin{figure}
\includegraphics{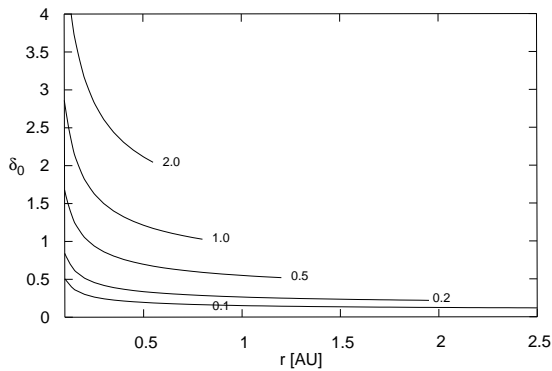}
\caption{The contours show the critical amplitude
$\delta_{k,\mathrm{crit}(\delta_0,r)}$ for which
$\nu(\delta_{k,\mathrm{crit}}) = \nu(0)$. Beyond their endpoints
the solution becomes unphysical ($\delta_k>\delta_0$, i.e. the
perturbed disk has regions with negative thickness).} \label{fig5}
\end{figure}

Initially, the function $\nu(\delta_k)$ always grows as $\delta_k$
increases. However, depending on the remaining parameters
($\delta_0$, $r$ and $l$), it may start to decrease and quickly
drop below the initial value $\nu(\delta_k = 0)$. Therefore, a
very small perturbation of the disc thickness is always stabilized
by the stellar irradiation, but if the amplitude of the
perturbation reaches some value, the ring instability may start to
work. This critical value $\delta_{k,\mathrm{crit}}$
($\nu(\delta_{k,\mathrm{crit}}) = \nu(0)$) strongly depends on
$l$: it is smaller for higher $l$, where the grazing angle is
smaller. The dependence on the remaining parameters ($r$ and
$\delta_0$) for $l=3$ is shown by Fig.~\ref{fig5}.

We see that the stabilizing effect of the irradiation may become
unimportant for broad rings (large $l$), small $\delta_0$ (less
mass accumulated in the dead zone) and/or at small radii.


\section{Discussion} 

We described a new instability in layered accretion discs. The
accretion rate in such discs is in general a function of radius.
As a result, mass may accumulate in the non-viscous dead zone near
the mid-plane of the disc. However, a small ring-like perturbation
of the dead zone thickness leads to the deviation of the
rotational velocity which results in non-uniform accumulation rate
of the mass supporting growing of that ring perturbation. This
ring instability may eventually lead to a decomposition of the
dead zone into rings.

We observed the formation of such rings in the 2D axially
symmetric radiation-hydrodynamic simulation of the layered disc.
To illustrate how the instability works, we give its approximate
analytical description. According to the analytical results, the
narrowest rings grow most rapidly. Therefore, we may expect the
formation of the radially thin rings whose radial extent will be
given just by the thermal pressure of the gas. The comparison of
ring sizes shows a reasonable agreement between the numerical
simulation and the analytical model.

The irradiation by the central star may substantially decelerate
or even stop the growth of the instability in some regions of the
disc. However, broad rings are less affected, and the effects of
irradiation become less important in thin discs and/or at small
distances from the star. On the other hand, once the innermost
ring has developed, and the flaring index $\gamma$ is not too
high, the disc at larger radii is shadowed and more rings may
develop there. If the disc is truncated (e.g. by the magnetosphere
of the star) the shadowing effect may also be caused by its inner
edge, allowing for an efficient growth of the instability.

The rings created by this instability may be important in terms of
planet formation, because they can be the places where the solid
particles (gravel and boulders) accumulate. It is because of the
higher rotational velocity at the inner edge of the ring and the
lower one at the outer edge. The drag force which makes the grains
to move inwards is smaller at the inner edge of the ring and
higher at the outer edge. The solid particles may merge into
larger bodies (planetesimals) necessary for the formation of
planets.

Massive rings are subject to a hydrodynamical instability in 3D,
e.g. with respect to non-axisymmetric perturbations
\citep{pp84,pp85}. This instability occurs if the rotational
angular velocity decreases with radius steeper than
$r^{-\sqrt{3}}$. In our simulation, this occurs for the innermost
ring at a time of $\sim 280$~yr. In the non-viscous case such ring
would most likely decay into so-called "planets", i.e. large scale
vortices, as found in numerical simulations by \citet{hawley87},
and further investigated by \citet*{gng87}. The fate of such a
ring in 3D viscous hydro-simulations including the effects of
layered accretion still have to be performed.

Our model assumes zero viscosity in the dead zone. However, there
are some indications \citep{fs03} that even in the dead one there
might be some turbulence present, induced by waves originating in
the active parts of the disc. In such a case, the excess of
surface density could result in a higher accretion rate in the
rings, and the growth of the rings would be stopped or even they
might not form at all. This issue will be addressed in a
forthcoming paper.


\section{Acknowledgments}

This research was supported in part by the European Community's Human
Potential Programme under contract HPRN-CT-2002-00308,PLANETS. RW
acknowledges financial support by the Grant Agency of the Academy of
Sciences of the Czech Republic under the grants AVOZ 10030501 and
B3003106. MR acknowledges the support from the grant No. 1 P03D 026 26
from the Polish Ministry of Science.


\label{lastpage}

\end{document}